\documentclass[amsmath,amssymb]{WileyMSP-template}
\usepackage{graphicx}% Include figure files
\usepackage{dcolumn}% Align table columns on decimal point
\usepackage{bm}% bold math
\usepackage{caption}
\usepackage{subcaption}

\usepackage{hyperref}% add hypertext capabilities
\usepackage[mathlines]{lineno}% Enable numbering of text and display math
\usepackage{amssymb, amsmath}
\usepackage{siunitx}

\usepackage[hang,flushmargin]{footmisc} 
\usepackage{ragged2e}
\justifying
\DeclareCaptionJustification{justified}{\justifying}
\newcommand\blfootnote[1]{%
  \begingroup
  \renewcommand\thefootnote{}\footnote{#1}%
  \addtocounter{footnote}{-1}%
  \endgroup
}

\begin{document}

\pagestyle{fancy}
%\rhead{\includegraphics[width=2.5cm]{vch-logo.png}}

\title{Tank-Circuit Assisted Coupling Method for Sympathetic Laser Cooling}

\maketitle

% Author: Please give full first and last names for authors and include * after the name of all corresponding authors

\author{Bingsheng Tu $^{1}$$^{,*,+}$}
\author{Felix Hahne $^{2}$$^{,+}$}
\author{Ioanna Arapoglou $^{1}$}
\author{Alexander Egl $^{1}$}
\author{Fabian Hei\ss e $^{1}$}
\author{Martin H{\"o}cker $^{1}$}
\author{Charlotte K{\"o}nig $^{1}$}
\author{Jonathan Morgner $^{1}$}
\author{Tim Sailer $^{1}$}
\author{Andreas Weigel $^{1}$}
\author{Robert Wolf $^{1}$$^{,\#}$}
\author{Sven Sturm $^{1}$}
\blfootnote{
\noindent Email Address: bingsheng.tu@mpi-hd.mpg.de\\
$^{+}$These authors have contributed to this work equally.\\
$^{\#}${Present address: ARC Centre for Engineered Quantum Systems, School of Physics, The University of Sydney, NSW 2006, Australia.}
}

% Dedication

\dedication{}

% Affiliations: Please provide adacemic titles (Prof. or Dr.) for all authors where applicable, and include an institutional email address for all corresponding authors
\begin{affiliations}

%\noindent Dr. B. Tu$^{[+]}$, F. Hahne$^{[+]}$, Dr. I. Arapoglou, Dr. A. Egl, Dr. F. Hei\ss e, Dr. M. H{\"o}cker, C. K{\"o}nig, J. Morgner, T. Sailer, Dr. A. Weigel, Dr. R. Wolf$^{[++]}$, and Dr. S. Sturm\\
\noindent $^{1}$Max-Planck-Institut f{\"u}r Kernphysik, Saupfercheckweg 1, 69117 Heidelberg, Germany\\
$^{2}$Fakult{\"a}t f{\"u}r Physik und Astronomie, Universit{\"a}t Heidelberg, 69120 Heidelberg, Germany\\
%\noindent  F. Hahne$^{[+]}$\\
\end{affiliations}

% Keywords: Please provide a minimum of three and a maximum of seven keywords, separated by commas


% Abstract should be written in the present tense and impersonal style (i.e., avoid we), and be at most 200 words long
\begin{abstract}

We discuss the coupling of the motion of two ion species in separate Penning traps via a common tank circuit. The enhancement of the coupling assisted by the tank circuit is demonstrated by an avoided crossing behavior measurement of the motional modes of two coupled ions. We propose an intermittent laser cooling method for sympathetic cooling and provide a theoretical description. The technique enables tuning of the coupling strength between two ion species in separate traps and thus allows for efficient sympathetic cooling of an arbitrary type of single ion for high-precision Penning-trap experiments.

\end{abstract}

% Text: Please use section headings and subheadings as specified below. For communications, all section headings apart from Experimental Section should be removed
% Please make the first reference to a display item bold: \textbf{Figure 1}
% Do not abbreviate Figure, Equation, etc.; display items are always singular, i.e., Figure 1 and 2.
% Equations are always singular, i.e., Equation 1 and 2, and should be inserted using the {equation} environment, not as graphics
% Please do not use footnotes in the text, additional information can be added to the Reference list.

\section{Introduction}
Penning traps have been proven as a versatile tool for fundamental physics \cite{Blaum2021,Myers2019} as well as in quantum science \cite{Jain2020,Nitzschke2020}. A multitude of high-precision measurements have been performed at Penning-trap facilities, such as measurements of atomic masses \cite{Sturm2014,Ulmer2015,Heisse2017,Eliseev2015,Rischka2020,Myers2020,Myers2015}, magnetic moments of elementary particles, for instance the free \cite{Gabrielse2008} and bound electron \cite{Sven2011,Arapoglou2019}, proton \cite{Mooser2014} and antiproton \cite{Smorra2017} and laser spectroscopy of a single highly charged ion (HCI) \cite{Egl2019}. These experiments are dedicated to testing pillars of the Standard Model of Physics, such as quantum electrodynamics (QED) \cite{Gabrielse2008,Sven2011,Arapoglou2019} and charge-parity-time (CPT) reversal symmetry \cite{Ulmer2015,Mooser2014, Smorra2017}, as well as to the determination of its parameters like the fine-structure constant $\alpha$ \cite{Gabrielse2008}. 

\noindent \textsc{Alphatrap} is such a Penning-trap experiment that allows performing precise $g$-factor measurements with HCIs, laser spectroscopy as well as mass measurements \cite{Sturm2019}. Two major projects to be carried out (or planned) at \textsc{Alphatrap} are the $g$-factor measurements of highly charged heavy ions up to $^{208}\textrm{Pb}^{81+}$, $^{208}\textrm{Pb}^{77+}$ and the precision laser spectroscopy of ro-vibrational transitions in single $\textrm{H}_2^+$. The $g$-factor measurement of highly charged lead ions will yield the most stringent test of bound-state QED theory in the strongest electric fields \cite{Shabaev2015}. The laser spectroscopy of ro-vibrational transitions of hydrogen molecular ions in turn will test QED theory and allow the determination of fundamental constants such as the proton-to-electron mass ratio and the Rydberg constant ($\textrm{R}_\infty$) to the 10 parts-per-trillion level and beyond \cite{Alighanbari2020}. If in the future the corresponding transitions can be measured in the antimatter equivalent $\bar{\textrm{H}}_2^-$ ion, the comparison yields a strong test of the CPT symmetry\cite{Myers2019}. 

\noindent At \textsc{Alphatrap} the motional temperature of ions is currently thermalized with a cryogenic superconducting tank circuit (resonator), which in turn is cooled by liquid helium to about \SI{4.2}{\kelvin}. Owing to the particle oscillation amplitude, the accuracy of frequency determinations is limited by systematic uncertainties caused by anharmonicities arising from field imperfections and eventually special relativity. In $g$-factor measurements, a lower particle temperature furthermore decreases axial frequency fluctuations and thus helps to achieve higher fidelity of the spin-state determination, which is specifically crucial for heavy ions with a comparably small magnetic moment, such as boronlike $^{208}\textrm{Pb}^{77+}$, and also for nuclear $g$-factor measurements of for example the (anti-)proton and the $^3\textrm{He}$ nucleus \cite{Smorra2015,Mooser2018}. For the laser spectroscopy of $\textrm{H}_2^+$, one direct advantage of a lower ion temperature is the reduction of the Doppler width and the possibility to ultimately reach the Lamb-Dicke regime \cite{Alighanbari2018}, which is key to resolving the transitions at the $10^{-11}$ level and beyond. To reduce the ion temperature into the millikelvin regime, one solution is to use laser cooling, a technique commonly applied in rf traps \cite{Monroe1995}. Other techniques such as cooling the tank circuit further with a dilution refrigerator are limited to somewhat higher temperatures and raise significant challenges for the cryomechanical design of experiments with external ion injection like the \textsc{Alphatrap} experiment. For direct laser cooling most of the ion species do not have suitable transitions. This is true for almost all HCIs, but also for many singly charged light ions like $\textrm{H}_2^+$. Hence, these species have to be cooled sympathetically by co-trapping auxiliary ions such as $^9\textrm{Be}^+$. This technique has been successfully implemented in rf traps for quantum logic spectroscopy and modern metrology \cite{Schmoger2015,Micke2020}. Sympathetic cooling in Penning traps has been studied already in 1986 when the Wineland group demonstrated cooling of a $^{198}\textrm{Hg}^+$ plasma by laser cooled $^9\textrm{Be}^+$ ions. Later that process was also shown with other species, e.g with Xe ions \cite{Evidence2001} and a positron plasma \cite{Jelenkovi2003}. However, such setups are not suitable for cooling a single, unperturbed ion to millikelvin temperatures. The co-trapping of different ions leads to drastic modifications of the motion of the ion of interest via the Coulomb interaction, which hinders precise determinations of the motional frequency as required for many high-precision Penning-trap experiments. To solve the Coulomb disturbance, a cooling technique proposed by Heinzen and Wineland is based on the coupling of two species in separate traps that share a common endcap electrode \cite{Wineland1990}. This way, the ions can interact via their image charges induced into the shared endcap electrode and the direct Coulomb interaction is suppressed due to the large distance. However, with realistic trap parameters \cite{Bohman2018} the coupling strength is too small to achieve highly efficient sympathetic cooling. 

To this end, we present a new coupling technique for two ion species in separate traps via a common tank circuit which can significantly enhance the coupling strength. After a description of the general common endcap coupling method (see Section \ref{comEC}), the theoretical basis of this method is presented (see Section \ref{prin}). A single $\textrm{H}_2^+$ ion and a cloud of $^9\textrm{Be}^+$ ions are used as examples in the calculation of the coupling parameters, owing to their importance for the planned laser spectroscopy project. For any HCI, such as $^{208}\textrm{Pb}^{81+}$, the principle of coupling is identical and it is easier to be implemented in practice due to the higher charge. In the second part of this paper we investigate the coupling experimentally by observing the resulting avoided crossing of the coupled frequencies. Since at the time of this work ALPHATRAP did not yet have a dedicated coupling trap suitable for $^9\textrm{Be}^+$ ions, this demonstration has been carried out using HCIs (see Section \ref{Exptalpha}). The results can however directly be transferred to other species. In section \ref{splc}, we propose a laser cooling scheme which consists of repetitive cycles of Doppler cooling and ion-ion coupling and provide simulations of the cooling of a singly charged $\textrm{H}_2^+$ ion and a highly charged $^{208}\textrm{Pb}^{81+}$ ion to the \SI{10}{\milli\kelvin} regime under realistic trap conditions.

\section{Common Endcap Coupling and Sympathetic Laser Cooling}\label{comEC}

In a Penning trap, the superposition of a strong homogeneous magnetic field and an electrostatic quadratic potential confines charged particles in all spatial dimensions. The motional trajectories of a single particle are the superposition of three harmonic oscillators: the fast and slow circular motions with the so-called modified cyclotron frequency ($\omega_+$) and the magnetron frequency ($\omega_-$) in the radial plane, and the axial harmonic oscillation with frequency $\omega_{z}$. In the case that two ions are coupled axially via a common endcap electrode, the axial equations of motion of the coupled ions have the form:
\begin{equation}
\begin{split}
&  \ddot{z_1} = -\widetilde{\omega}_1^2z_1-C_{12}z_2-\gamma_{11}\dot{z_1}-\gamma_{12}\dot{z_2}-\frac{q_1}{m_1D_1}U_\textrm{noise},\\
&  \ddot{z_2} = -\widetilde{\omega}_2^2z_2-C_{21}z_1-\gamma_{22}\dot{z_2}-\gamma_{21}\dot{z_1}-\frac{q_2}{m_2D_2}U_\textrm{noise}.
\end{split}
\label{EOM1}
\end{equation}
\noindent Here, $C_{ij} = \frac{N_{j}q_{i}q_{j}}{m_{i}D_{i}D_{j}C_\textrm{T}}$, where, $i$ and $j$ denote the two ion species, $N_{i}$ is the number of ions of type $i$, $q_{i}$, $m_{i}$ are the charge and mass, respectively. $z_i$ is the axial position and $\widetilde{\omega}_i$ is the effective axial frequency (including the shift resulting from the self-interaction via the endcap). $C_\textrm{T}$ and $D_{i}$ are the equivalent parallel capacitance and the effective distance \cite{Bohman2018} of the common endcap. ${U_\textrm{noise}}$ denotes a noise excitation (typically the Johnson thermal noise of a resonator) and $\gamma_{ij}$ represents a damping term due to either laser cooling or resistive cooling from a tank circuit (see section below). Equation (\ref{EOM1}) is a second order differential equation for which the solution has a fast rotating component (the axial oscillation with frequency $\widetilde{\omega_{i}}$) and a slow rotating component (the axial energy exchange between the two ion species). A rotating wave approximation (RWA) can be employed to simplify Equation (\ref{EOM1}) by defining ${z_{i}} \equiv \frac{A_{i}}{2}e^{i\widetilde{\omega_{i}}t}+\frac{A_{i}^{*}}{2}e^{-i\widetilde{\omega_{i}}t}$ with the time-dependent axial amplitude $A_{i}$. After eliminating the counter-rotating terms, we get the form:

\begin{equation}
\begin{split}
&  \dot{A_1} = \frac{i}{2\widetilde{\omega}_{1}}(C_{12}A_2e^{-i\Delta\omega t}+i\widetilde{\omega}_{1}\gamma_{11}A_1+i\widetilde{\omega}_{2}\gamma_{12}A_2e^{-i\Delta\omega t}-\frac{2q_1}{m_1D_1}U_\textrm{noise}),\\
&  \dot{A_2} = \frac{i}{2\widetilde{\omega}_{2}}(C_{21}A_1e^{i\Delta\omega t}+i\widetilde{\omega}_{2}\gamma_{22}A_2+i\widetilde{\omega}_{1}\gamma_{21}A_1e^{i\Delta\omega t}-\frac{2q_2}{m_2D_2}U_\textrm{noise}).
\end{split}
\label{RWA1}
\end{equation}

\noindent Here, $\Delta\omega = \widetilde{\omega}_{1}-\widetilde{\omega}_{2}$. In the case of exact resonance ($\widetilde{\omega}_{1} = \widetilde{\omega}_{2} = \widetilde{\omega}$) and in absence of damping and noise excitation, the solution has the form ${A_{i}}(t) = \frac{A_{i0}}{2}(e^{i(\frac{\Omega_\textrm{R}}{2}t+\phi_{i0})}+e^{-i(\frac{\Omega_\textrm{R}}{2}t+\phi_{i0})})$, where

\begin{equation}
\Omega_\textrm{R} = \frac{\pi}{\tau_\textrm{ex}}= \frac{q_1q_2}{\widetilde{\omega} D_1D_2C_\textrm{T}}\sqrt{\frac{N_1N_2}{m_1m_2}}.
\label{Rabi}
\end{equation}

\noindent Here, $\tau_\textrm{ex}$ is the energy exchange time in the coupled system and $\phi_{i0}$ denotes the initial phase.

\begin{figure}[!h]
%\begin{center}
\centering
\includegraphics[width=0.5\columnwidth]{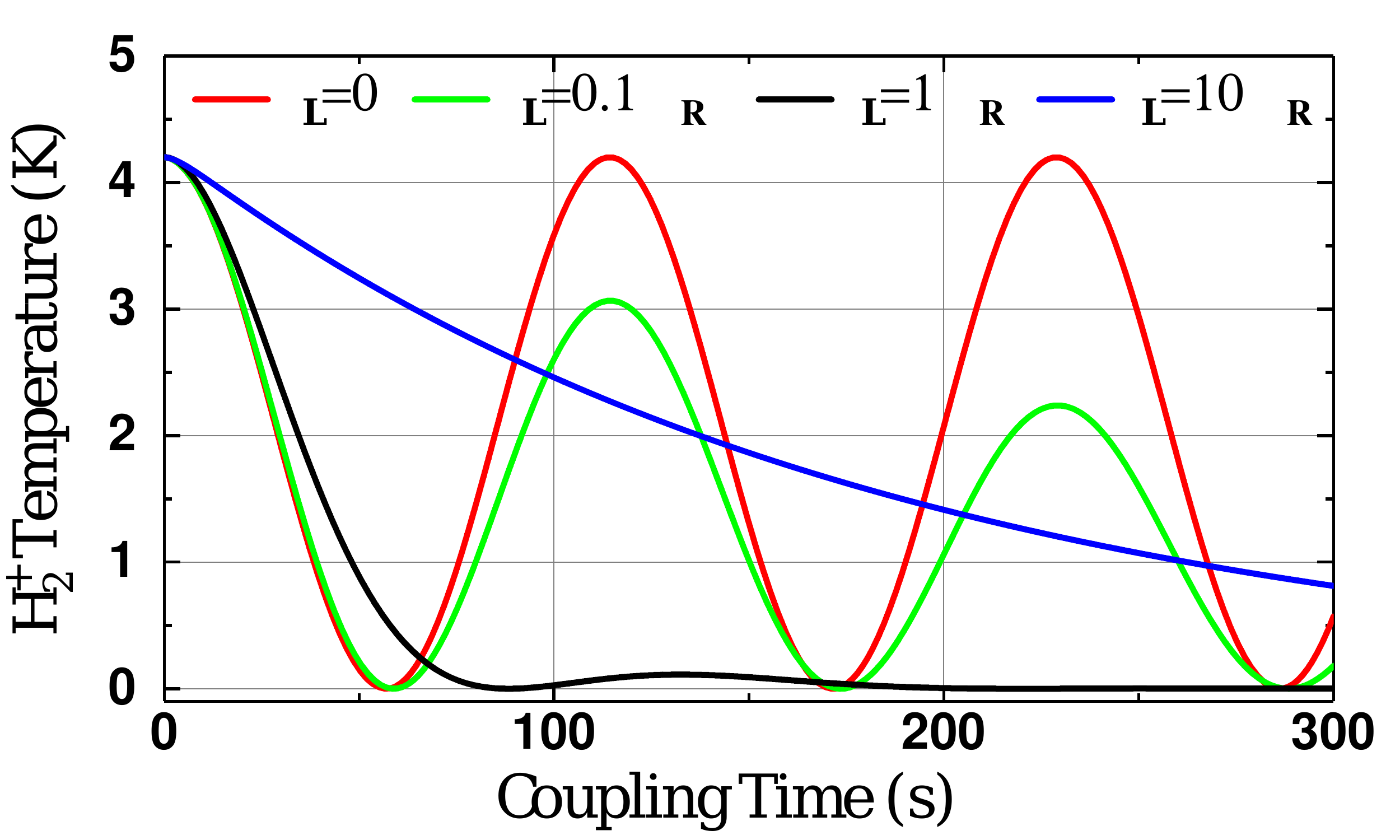}
\caption[]{The sympathetic cooling of a single $\textrm{H}_2^+$ ion with 100 $^9\textrm{Be}^+$ ions via a common endcap with different laser damping coefficients $\gamma_\textrm{L}$. $\Omega_\textrm{R}$ is the Rabi frequency given in Equation (\ref{Rabi}). For details see text.}
%\end{center}
\label{Couple} 
\end{figure}

\noindent In this ideal case, the two coupled ion species can exchange their respective energy with the Rabi frequency $\Omega_\textrm{R}$, which is an analog to the Rabi oscillation of a two-level system. With a realistic coupling trap geometry ($C_\textrm{T} = $ \SI{10}{\pico\farad}, $D_1 = D_2 = $ \SI{4.6}{\milli\meter} and $\widetilde{\omega} = 2\pi \times 500$ \SI{}{\kilo\hertz}, see ref. \cite{Bohman2018}), we have numerically simulated coupling of a single $\textrm{H}_2^+$ ion (ion 1) with a cloud of $^9\textrm{Be}^+$ ions (ion 2, $N=100$) (shown in Figure \ref{Couple}). Initially, the $\textrm{H}_2^+$ ion is thermalized to the environmental temperature $T_0 = $ \SI{4.2}{\kelvin} while the center-of-mass motion of the cloud of $^9\textrm{Be}^+$ ions is pre-cooled by the laser to the Doppler limit $T_\textrm{D} \approx $ \SI{0.5}{\milli\kelvin}. The axial motion of the single $\textrm{H}_2^+$ ion and the center-of-mass motion of the cloud of $^9\textrm{Be}^+$ ions can be coupled when their axial frequencies are sufficiently close. Without laser cooling, the two species periodically exchange their energies with $\tau_\textrm{ex} = $ \SI{57}{\second} (see red curve in Figure \ref{Couple}). With laser cooling of the $^9\textrm{Be}^+$ ions, treated here as a damping $\gamma_\textrm{22} = \gamma_\textrm{L}$ in Equation (\ref{EOM1}), the single $\textrm{H}_2^+$ ion can be sympathetically cooled to $T_\textrm{D}$. Following the cooling of the axial mode, axial-to-radial radiofrequency sideband drives can be used to cool all dimensions of motion.

\begin{figure}[!h]
%\begin{center}
\centering
\includegraphics[width=0.5\columnwidth]{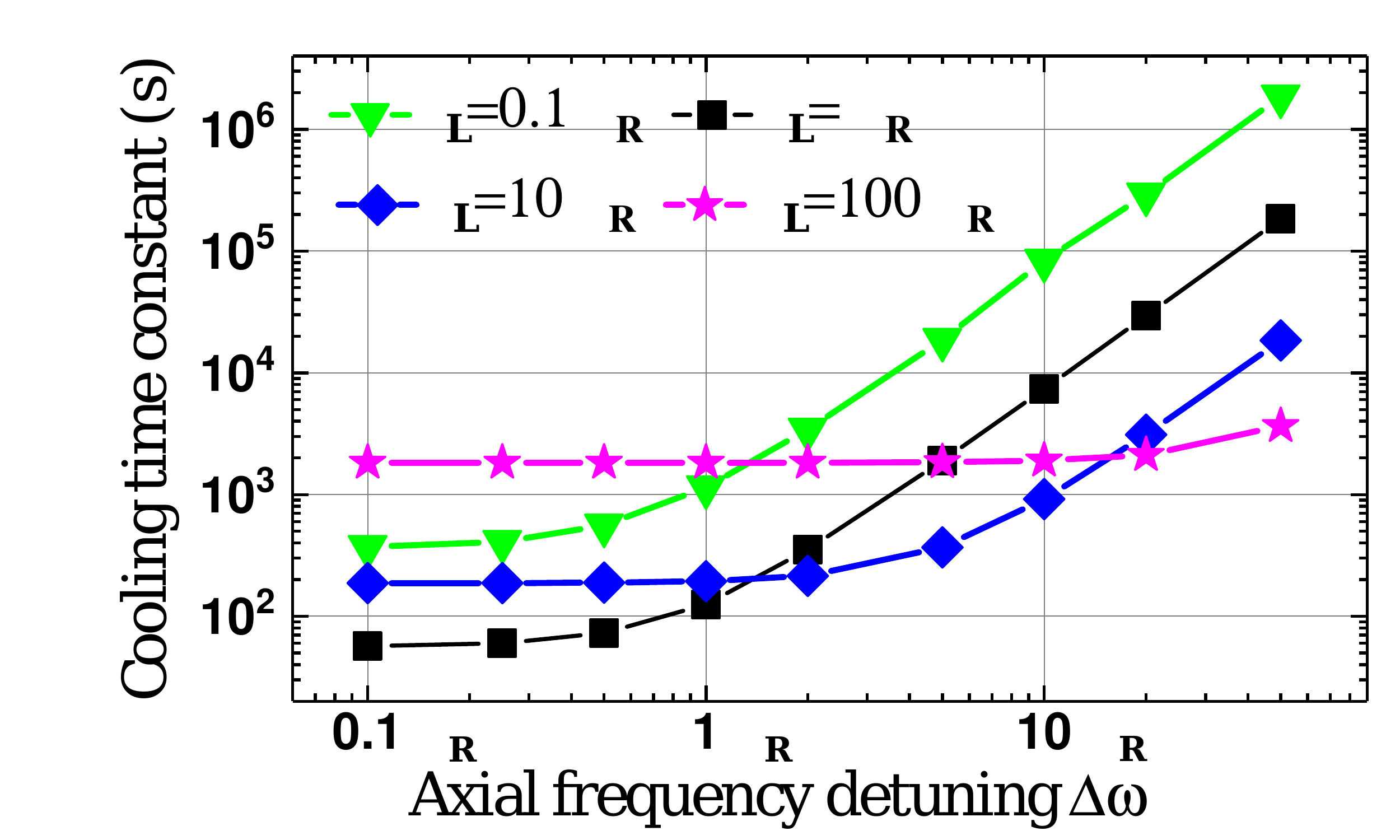}
\caption[]{The calculated cooling time constant $\tau_\textrm{cool}$ as a function of axial frequency detuning $\Delta{\omega} = \widetilde{\omega_1}-\widetilde{\omega_2}$ with different laser damping coefficients $\gamma_\textrm{L}$.}
%\end{center}
\label{LaserOn} 
\end{figure}

\noindent As expected from a system of weakly coupled oscillators, Figure \ref{Couple} shows different effective cooling rates depending on the laser damping coefficient. We define a cooling time constant $\tau_\textrm{cool}$ where the total axial energy of the system has reduced by a factor of $e$. The more general dependency of $\tau_\textrm{cool}$ on both the laser damping coefficient and the axial frequency detuning ($\Delta{\omega} = \widetilde{\omega}_{1}-\widetilde{\omega}_{2}$) is shown in Figure \ref{LaserOn}. It can be seen that the sympathetic cooling becomes maximally effective when the damping coefficient coincides with the Rabi frequency $\Omega_\textrm{R}$ which is about $ 2\pi \times 9$ \SI{}{\milli\hertz} in this case. This is because for faster laser cooling rates the Rabi oscillation becomes overdamped as the motion of the $^9\textrm{Be}^+$ ions is cooled faster than energy can be transferred from ion 1. In this regime the cooling rate of ion 1 decreases for stronger laser cooling rates because the motion of the $^9\textrm{Be}^+$ ions becomes spectrally broadened by the damping and the energy transfer is consequently less efficient. In the extreme case of laser cooling rates in the MHz regime, the motion essentially comes to a rest and the cooling of the ion of interest stops. Reaching the optimal damping coefficient requires fine-tuning of the laser power and frequency offset. Furthermore, if the two axial frequencies are not exactly identical (compared to $\Omega_\textrm{R}$), the efficiency of the cooling also decreases as now only a fraction of the energy is exchanged. This becomes an issue especially for low Rabi frequencies, as the control and stability of the axial frequencies is typically limited by the stability of the voltage sources (UM1-14 \cite{um114}, StaRep \cite{BOHM2016}) to 30 ppb relative or about 15mHz absolute at best.

\noindent In order to relax the experimental constraints, the coupling strength should be enhanced. In this work, we propose a new coupling method with the assistance of a parallel resonant tank circuit. This way, by shunting a fraction of the trap capacitance via the common resonator, a much larger and also tunable coupling impedance, equivalent to a very small effective capacitance in Equation (\ref{Rabi}), can be achieved  and thus the coupling of the motion of the two ion species is significantly enhanced. Additionally, we propose an optimized laser cooling scheme that allows achieving efficient sympathetic cooling without the need to fine-tune the laser power or detuning. Another advantage of the common tank circuit coupling is that the two traps can be spatially separated, which provides convenience for operation and precision measurements of the ion of interest of arbitrary type and charge in an isolated trap with optimized field geometry.

\section{Tank-Circuit Assisted Coupling Method}\label{TCACM}

\subsection{Theory}\label{prin}

A resonant tank circuit that is employed as a resonator for the detection of the image current signals of the trapped ions typically consists of a superconducting coil and the combined capacitance of the coil ($C_\textrm{R}$), the wiring and the trap ($C_\textrm{T}$). The total impedance $Z_\textrm{LC}$ of the resonator has the form: 
\begin{equation}
%\begin{split}
 Z_\textrm{LC} = \left(\frac{1}{R_\textrm{p}}+i\omega C+\frac{1}{i\omega L}\right)^{-1} =\frac{R_\textrm{p}}{\left[1+iQ(\frac{\omega}{\omega_\textrm{R}}-\frac{\omega_\textrm{R}}{\omega})\right]}.
\label{res}
%\end{split}
\end{equation}
Here, ${R_\textrm{p}}$ is the equivalent parallel resistance, which represents all losses of the circuit. $C$ and $L$ denote the parallel capacitance and inductance, respectively. $Q=\frac{\omega_{\textrm{R}}}{\textrm{FWHM}}$ is the quality factor, FWHM is the full width half maximum, and $\omega_\textrm{R}$ is the resonance frequency of the tank circuit. For the explanation of the new coupling method, the total impedance is expressed in terms of a real part and an imaginary part: 
\begin{equation}
\begin{split}
 & \textrm{Re}(Z_\textrm{LC}) =\frac{R_\textrm{p}}{1+Q^2\left(\frac{\omega_\textrm{R}+\textrm{d}\omega}{\omega_\textrm{R}}-\frac{\omega_\textrm{R}}{\omega_\textrm{R}+\textrm{d}\omega}\right)^2} \\
& \textrm{Im}(Z_\textrm{LC}) =-\frac{R_\textrm{p}Q\left(\frac{\omega_\textrm{R}+\textrm{d}\omega}{\omega_\textrm{R}}-\frac{\omega_\textrm{R}}{\omega_\textrm{R}+\textrm{d}\omega}\right)}{1+Q^2\left(\frac{\omega_\textrm{R}+\textrm{d}\omega}{\omega_\textrm{R}}-\frac{\omega_\textrm{R}}{\omega_\textrm{R}+\textrm{d}\omega}\right)^2}.\\
\end{split}
\end{equation}
Here, $\textrm{d}\omega$ is the frequency detuning from the resonator frequency $\textrm{d}\omega = \omega - \omega_\textrm{R}$. When two ion species are coupled via a resonant tank circuit, the impedance of that resonator determines the coupling strength. In resonance $\textrm{d}\omega = 0$ the real part of the total impedance (resistance) reaches its maximum $R_\textrm{p}$, while the imaginary part vanishes. Above the resonance frequency, with a frequency detuning larger than the resonance width (the typical experimental condition) the real part drastically drops and the impedance is dominated by the imaginary part. We can then accurately model the general impedance with an effective resistance $R_\textrm{eff} \equiv \textrm{Re}(Z_\textrm{LC}) \approx \frac{R_\textrm{p}(\omega_\textrm{R}+ \textrm{d}\omega)^2}{4Q^2\textrm{d}\omega^2}$ and an effective capacitance $C_\textrm{eff} \equiv -\frac{1}{\textrm{Im}(Z_\textrm{LC})\omega} \approx \frac{2C  \omega_\textrm{R} \textrm{d}\omega}{(\omega_\textrm{R}+d \omega)^2}$, which determines the coupling strength. The effective resistance results in dissipation for hot ions and a heating rate for ions that are colder than the (typically) \SI{4.2}{\kelvin} environment. The strength of this coupling is expressed as $\gamma_{i-\textrm{Res.}}=\frac{1}{\tau_{i-\textrm{Res.}}}=\frac{N_{i}q_{i}^2R_{\textrm{eff}}}{m_{i}D_{i}^2}$.

\begin{figure}[!h]
%\begin{center}
\centering
\includegraphics[width=0.5\columnwidth]{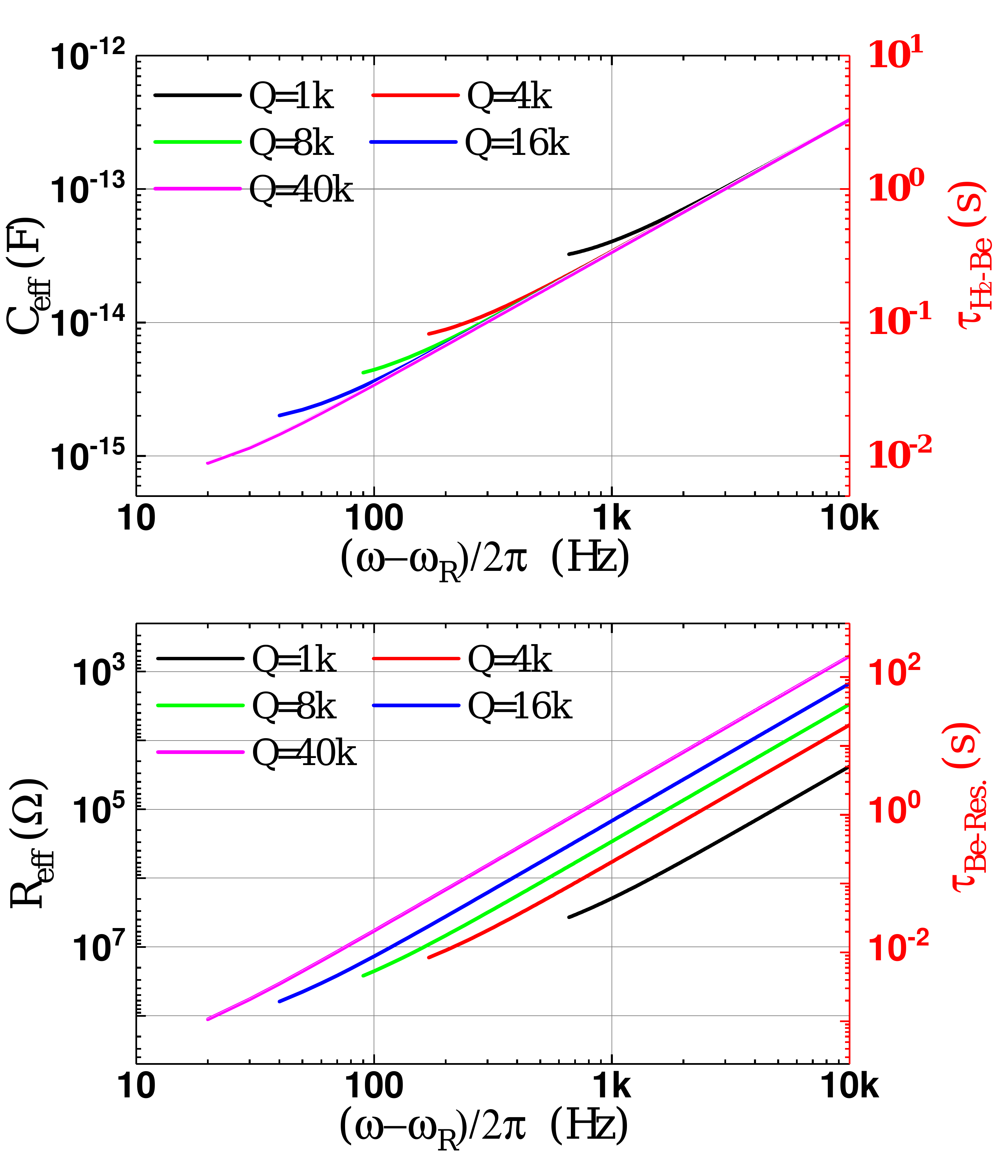}
\caption[]{The calculated effective capacitance $C_\textrm{eff}$ (upper plot) and resistances $R_\textrm{eff}$ (lower plot) of the resonant tank circuits with different quality factors $Q$ as a function of frequency detuning $ (\omega - \omega_\textrm{R} )/ 2 \pi $. The plots only show $C_\textrm{eff}$ and $R_\textrm{eff}$ for each resonator with a frequency detuning larger than the resonance width. As a function of detuning, $R_\textrm{eff}$ decreases quadratically while $C_\textrm{eff}$ (inverse imaginary impedance) increases linearly. As an example, the $\textrm{H}_2$-$\textrm{Be}$ coupling time constant and $\textrm{Be}$-resonator coupling time constant are shown on the right vertical axes. }
%\end{center}
\label{Zeff} 
\end{figure}

\noindent As a realistic example we can assume a resonant tank circuit as used in \textsc{Alphatrap} ($L = $ \SI{2.1}{\milli\henry}, $C_\textrm{R} = $ \SI{5.1}{\pico\farad}) to be connected to a common endcap ($C_\textrm{T} = $ \SI{10}{\pico\farad}) to assist the $\textrm{H}_2 - ($N=100$) \textrm{Be}$ coupling. In Figure \ref{Zeff} the effective capacitance $C_\textrm{eff}$ and resistance $R_\textrm{eff}$ are calculated and plotted as a function of frequency detuning $(\omega - \omega_\textrm{R})/ 2\pi$ for different $Q$-values. The effective capacitance (upper plot) can be significantly reduced by orders of magnitude as compared to $C_\textrm{T}$, which leads to a drastic enhancement of the ion-ion coupling as shown on the right axis scale. Additionally, with a suitable frequency detuning and a high-$Q$ resonator, the ion-resonator coupling (lower plot) can be reduced to such an extent that the ion-resonator coupling strength will not overpower the ion-ion coupling (comparing both time constants on the right axis scale). In the case of sympathetic cooling, owing to the heating effect of the tank circuit, the final equilibrium temperature of the system is always higher than the Doppler limit. Nevertheless, the enhancement of ion-ion coupling significantly relaxes the experimental constraints as discussed in section \ref{comEC}. Additionally, the much shorter cooling time could enable a lower equilibrium temperature than the pure common endcap coupling if there is any noise near the motional frequency on the trap electrode. In section \ref{splc}, we will show how our proposed intermittent laser cooling technique can help to reduce the resonator heating and reach an optimum temperature with given trap conditions.

\subsection{Experiment}\label{Exptalpha}

\begin{figure}[!h]
%\begin{center}
\centering
\includegraphics[width=0.9\columnwidth]{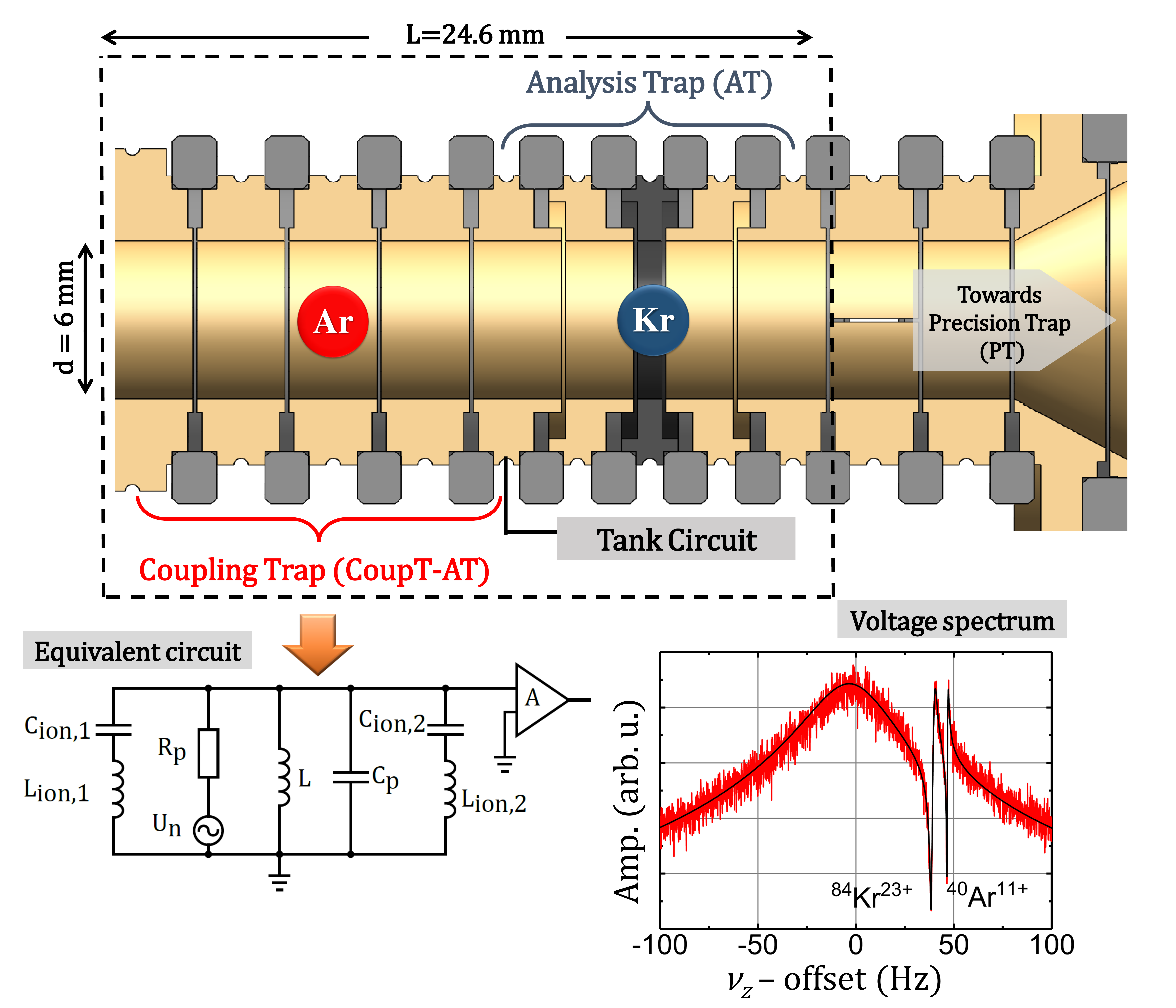}
\caption[]{Schematic of the experimental setup for ion-ion coupling via a common axial resonator. The axial motions of a single $^{84}\textrm{Kr}^{23+}$ and a single $^{40}\textrm{Ar}^{11+}$ ion, which are located in the AT and CoupT-AT, respectively, are detected, showing the “dip” signals on the wing of the common resonator.}
%\end{center}
\label{Trap} 
\end{figure}

\textsc{Alphatrap} has two cylindrical Penning traps, the precision trap (PT) and the analysis trap (AT). The AT (shown in Figure \ref{Trap}) has a ferromagnetic ring electrode that produces a magnetic bottle for spin-state detection and a medium-$Q$ axial  resonator ($R_\textrm{p} = 155$ \SI{}{\mega\ohm}, $L = $ \SI{10.5}{\milli\henry}, $C_\textrm{R} = $ \SI{6.3}{\pico\farad}, $C_\textrm{T} \approx $ \SI{15.2}{\pico\farad} and $Q\approx 7$k), while the PT (not shown) with a high-$Q$ axial resonator ($R_\textrm{p} = 344$ \SI{}{\mega\ohm}, $L = $ \SI{2.1}{\milli\henry}, $C_\textrm{R} = $ \SI{5.1}{\pico\farad}, $C_\textrm{T} \approx $ \SI{23.3}{\pico\farad} and $Q\approx 40$k) aims for precision measurements of ion frequencies. At the time of this work, a trap optimized to support the coupling of $\textrm{H}_2^+$ ion and $^9\textrm{Be}^+$ ions did not exist. Nevertheless, the principle of common tank-circuit coupling can be demonstrated by any other ion species. Some spare electrodes close to the AT were used to make a provisional coupling trap (CoupT-AT) (see Figure \ref{Trap}). The two HCIs, a single $^{84}\textrm{Kr}^{23+}$ and a single $^{40}\textrm{Ar}^{11+}$ ion, which were produced in a room-temperature compact electron beam ion trap \cite{Micke2018}, are trapped in the AT and CoupT-AT, respectively, with a compensated (harmonic to 4th order) potential setting. The common AT axial resonator ($\omega_\textrm{R}\approx 2\pi \times 334$ \SI{}{\kilo\hertz}), is connected to one of the electrodes between the two traps to assist the coupling. The effective electrode distances of both trap are $D_\textrm{AT}=19$ \SI{}{\milli\meter} and $D_\textrm{CoupT-AT}=25$ \SI{}{\milli\meter}, respectively. The very weak induced current signal due to the motion of the ions (a few \SI{}{\femto\ampere}) through the common resonator ($R_\textrm{p} = 155$ \SI{}{\mega\ohm}) is then amplified for read out. The spectrum (see inset in Figure \ref{Trap}) shows the typical “dip” signal of two ions once they are in thermal equilibrium with the Johnson thermal noise of the resonator. Generally, on the resonator the minimum position of the dip indicates the ion frequency. However, in order to reduce the individual resonator-ion couplings, the axial frequencies should be detuned from $\omega_\textrm{R}$. In this way, the axial frequencies are modified by the resonator's effective capacitance to $\widetilde{\omega}_{i} \approx \omega_{i} + \frac{q_{i}^2}{2m_{i} \omega_{i} D_{i}^2 C_{\textrm{eff}}}$. Once the two axial frequencies $\widetilde{\omega}_{i}$ on the wing of the resonator are close to each other, the ions become coupled and start exchanging energy. 

\noindent For the coupled ion system discussed in section \ref{comEC}, alternative orthogonal motional modes can be defined: the common motion mode $u=z_1+\alpha z_2$ and the counter motion mode  $v=z_1-\beta z_2$, where $\alpha$ and $\beta$ are the coefficients that depend on the ion species and trap geometry. The oscillation frequencies of the two modes are $\omega_u=\frac{1}{2}(\widetilde{\omega}_{1} + \widetilde{\omega}_{2}+ \sqrt{\Delta\omega^2 + \Omega_\textrm{R}^2})$ and $\omega_v=\frac{1}{2}(\widetilde{\omega}_{1} + \widetilde{\omega}_{2} - \sqrt{\Delta\omega^2 + \Omega_\textrm{R}^2})$. Figure 5a is a simulation of these two orthogonal modes where we set $^{40}\textrm{Ar}^{11+}$ as ion 1 and $^{84}\textrm{Kr}^{23+}$ as ion 2. The two-dimensional plot is the voltage spectrum (in arbitrary decibel units) as a function of the ions’ frequency difference $\Delta\nu = \Delta\omega/2\pi$ and the overall frequency detuning from the resonance frequency $(\omega - \omega_R)/2\pi$. In the 2D plot, each column of data is the spectrum of two ions (like the inset in Figure \ref{Trap}) in a small range about \SI{149.5}{\hertz} off the resonance frequency of the tank circuit. The x-axis indicates the frequency of ion 1 swept linearly across the fixed frequency of ion 2 by $\pm\SI{3}{\hertz}$ by adjusting the voltage sources. Assuming no interaction between the two ions, a simple crossing of the two ion signals is expected. However, the coupling gives rise to an avoided crossing behavior and the new orthogonal modes $u$ and $v$ belong to the upper and lower peak signal curves. The minimal splitting is given by the Rabi frequency $\Omega_\textrm{R}/2\pi$ at $\Delta\omega=0 \: (\widetilde{\omega}_{1}=\widetilde{\omega}_{2})$, where the coupling strength is maximal. With the frequency detuning $\Delta\omega$, the modified Rabi frequency can be derived as $\Omega_\textrm{R} ^{\prime}=\sqrt{\Omega_\textrm{R}^2+\Delta\omega^2}$. 
 
\begin{figure}
\centering
\begin{subfigure}[!h]{0.45\columnwidth}
   \includegraphics[width=1\columnwidth]{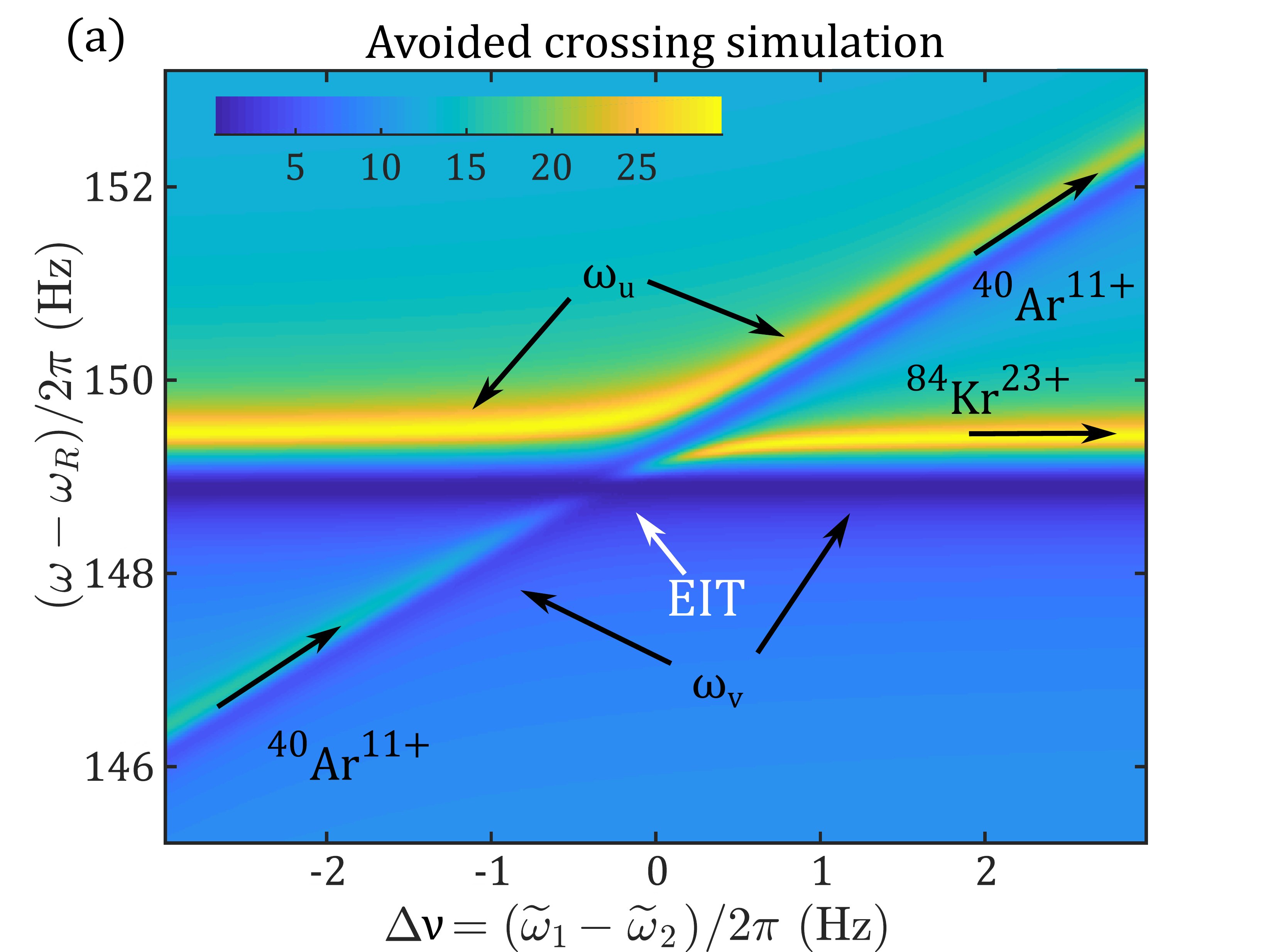}
   \label{avsim} 
\end{subfigure}

\begin{subfigure}[!h]{0.45\columnwidth}
   \includegraphics[width=1\columnwidth]{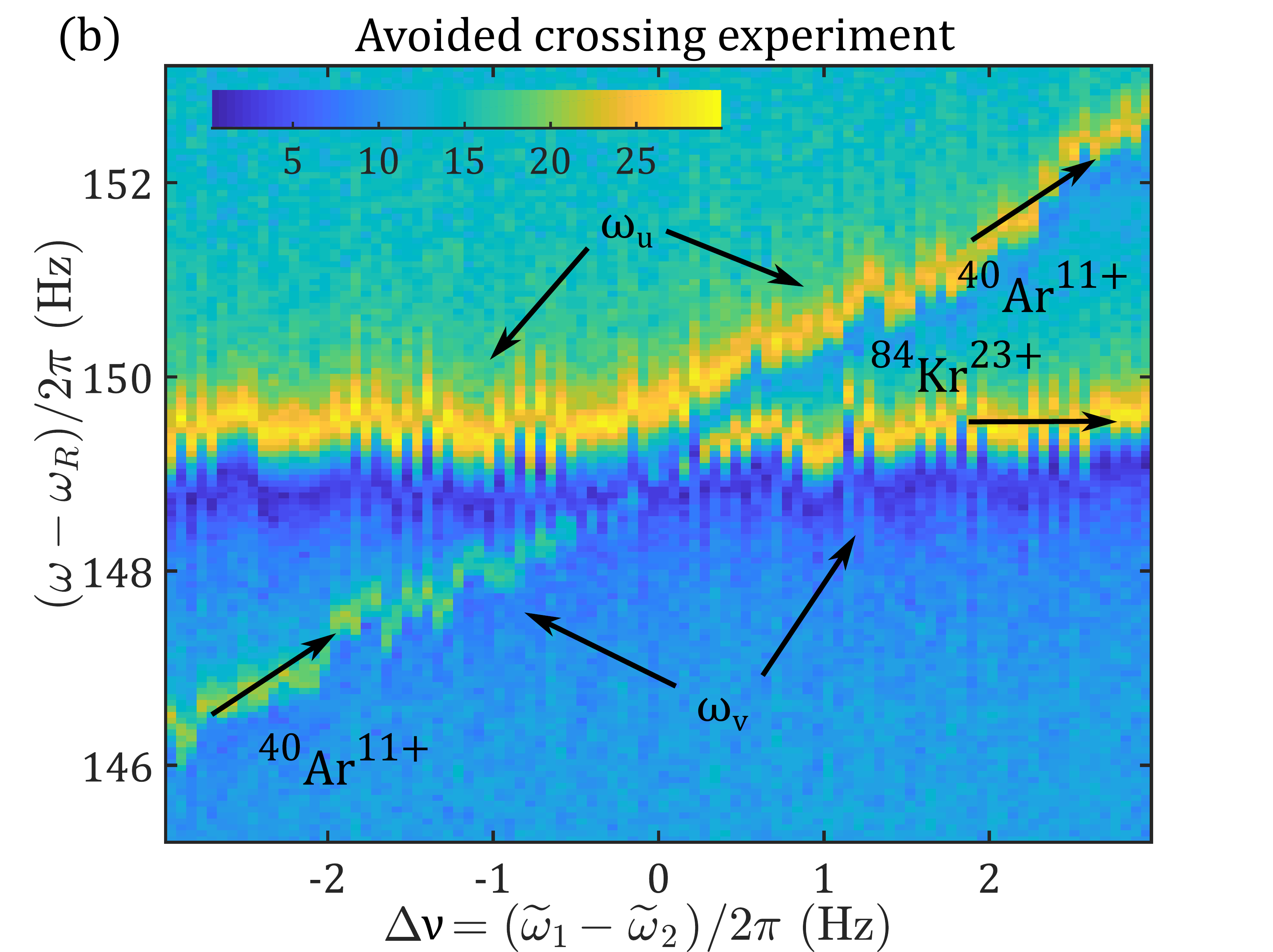}
   \label{avexp}
\end{subfigure}
\caption{The simulated (a) and measured (b) two-dimensional avoided crossing spectrum of the $^{40}\textrm{Ar}^{11+}$ ion and the $^{84}\textrm{Kr}^{23+}$ ion, as a function of the frequency difference $\Delta\nu$ and overall frequency detuning from the resonance frequency $ (\omega - \omega_\textrm{R} )/ 2 \pi $. The signal strength is given by the color coded bar (in arbitrary logarithmic units).}
\end{figure}

\noindent The experimental demonstration of the avoided crossing behavior is shown in Figure 5b. A similar 2D lineshape model \footnote {In the fitting procedure, a linear approximation of the effective capacitance has been used based on $C_\textrm{eff} \approx \frac{2C  \omega_\textrm{R} \textrm{d}\omega}{(\omega_\textrm{R}+d \omega)^2}$  with ion frequency scanning by $\pm\SI{3}{\hertz}$. This approximation is reasonable according to the calculation of this resonator model and the discrepancy induced into the result is negligible. } as used in the simulation was used to fit the spectra. From the fit results we deduce an effective capacitance $C_{\textrm{eff}}=1.982(2)_{\textrm{stat}}(_{-35}^{+39})_{\textrm{sys}} \times 10^{-14}$ \SI{}{\farad}, corresponding to a Rabi frequency $\Omega_R = 2\pi\times\SI{0.577}{\hertz}$, which agrees with the calculated (see derivations in section \ref{prin}) $C_{\textrm{eff,cal}}=1.978 \times 10^{-14}$ \SI{}{\farad} at \SI{149.5}{\hertz} offset to the resonator. This value is a factor of 760 smaller than the nominal electrode capacitance $C_\textrm{T,AT} = 15.2$ \SI{}{\pico\farad} leading to an enhancement of the coupling strength by the same amount. The major systematic uncertainty comes from the determination of ion frequencies and effective electrode distances of the AT and CoupT-AT. 

\noindent Notice the vanishing of the counter mode $v$ in the avoided crossing spectrum, indicated by the white marker in Figure 5a. The two ions oscillate such that their respective induced currents cancel each other in this mode, making the counter motion mode transparent for the resonator. This is a direct analog to electromagnetically induced transparency (EIT) \cite{Harris1997}. In the equivalent circuit (see inset in Figure \ref{Trap}), the induced current of this mode directly flows between the two ions, making the mode ``invisible" to the resonator and its noise heating. This ``invisibility" also implies no energy exchange with the tank circuit, which principally enables cooling of this mode to extremely low temperatures without any heating effect.

\section{Intermittent Laser Cooling}\label{splc}
In the intermittent laser cooling method, instead of continuously laser cooling with a critical power or frequency detuning (see discussion in Section \ref{comEC}), we plan to use a repetitive sequence of consecutive laser cooling and coupling cycles. Each cycle consists of Doppler cooling of $^9\textrm{Be}^+$ ions within some \SI{10}{\micro\second}, which is much shorter than the Rabi cycle but long compared to the axial motion and an (no-laser) interval $\tau_{\textrm{c}}$ for ion-ion coupling and energy transfer. This way, the auxiliary $^9\textrm{Be}^+$ ions can be initially cooled to $T_\textrm{D}$ and after the time interval $\tau_{\textrm{c}}$ which is shorter than the nominal $\tau_{\textrm{ex}}$ a fraction of the energy of the target ion (typically a few tens of millikelvin depending on $\tau_{\textrm{c}}$) is transferred to the auxiliary ion. This excess energy of the auxiliary ion is again cooled by the (saturated) laser beam almost instantly so that the coupling is re-initialized with the $^9\textrm{Be}^+$ at $T_\textrm{D}$ and the target ion at a few tens of millikelvin lower than before. By repeating the laser cooling and coupling multiple times, the target ion can be sympathetically cooled with no need for any de-coupling during the whole cooling period. In this way, the coherence in the coupling of the two axial motions is required only for $\tau_{\textrm{c}}$ rather than the longer $\tau_{\textrm{ex}}$, which relaxes one of the experimental constraints. Furthermore, the heating due to the ion-resonator coupling as well as possible excess noise can also be reduced, leading to a lower equilibrium temperature. In comparison with the continuous laser cooling method (we will show in the end of this section), the intermittent laser cooling is more robust because no fine tuning of the damping constant of the laser cooling is required. 

\noindent An analytical solution for this intermittent cooling technique is presented in this work. In one coupling time period $\tau_{\textrm{c}}$, the energy change for ion 1 can be calculated:
\begin{equation}
\Delta E_1(\tau_\textrm{c}) = -{\left\langle \Delta E_{1 \rightarrow 2,\tau_\textrm{c}} \right\rangle}+{\left\langle \Delta E_{2 \rightarrow 1,\tau_\textrm{c}} \right\rangle}+k_\textrm{b}T_0\frac{\tau_\textrm{c}}{\tau_\textrm{1-Res.}}.
\label{DelE1}
\end{equation}

\noindent Here, the subscripts 1 and 2 represent the target ion and the laser cooled ions, respectively, which are $\textrm{H}_2^+$ and $^9\textrm{Be}^+$ ions in this case. The term $k_\textrm{b}T_0\frac{\tau_\textrm{c}}{\tau_\textrm{1-Res.}}$ represents a source of noise heating towards the tank circuit temperature $T_0$. ${\left\langle \Delta E_{1 \rightarrow 2,\tau_\textrm{c}} \right\rangle}$ and ${\left\langle \Delta E_{2 \rightarrow 1,\tau_\textrm{c}} \right\rangle}$ are the effective energies transferred from ion 1 to ion 2 and ion 2 to ion 1, respectively. In a coherent Rabi oscillation of a two-level system without any heating, ${\left\langle \Delta E_{{i} \rightarrow {j},\tau_\textrm{c}} \right\rangle} = E_{i} \sin^2(\frac{\Omega_\textrm{R} \tau_\textrm{c}}{2})$, where $E_{i}$ is the energy of ion i. However, in practice a heating from Johnson and excess noise cannot be neglected, especially for low temperatures close to $T_\textrm{D}$.  Because of the stochastic nature of this heating, the energy gain and transfer between the two ions is incoherent. Generally, a Wiener process \cite{Bernt2003} is expected under the effect of noise excitation on one ion, and the effective energies transferred can be calculated by using the It$\hat{\textrm{o}}$ isometry \cite{Bernt2003} under the condition  $\tau_\textrm{c} \ll \frac{\pi}{\Omega_\textrm{R}} \ll \tau_{i-\textrm{Res.}}$ : 
\begin{equation}
\begin{split}
& {\left\langle \Delta E_{1 \rightarrow 2,\tau_\textrm{c}} \right\rangle}=\frac{1}{4}\Omega_\textrm{R}^2 \tau_\textrm{c}^2\left(E_1(0)+k_\textrm{b}T_0\frac{\tau_\textrm{c}}{3\tau_\textrm{1-Res.}}\right), \\
& {\left\langle \Delta E_{2 \rightarrow 1,\tau_\textrm{c}} \right\rangle}=\frac{1}{4}\Omega_\textrm{R}^2 \tau_\textrm{c}^2\left(k_\textrm{b}T_\textrm{D}+k_\textrm{b}T_0\frac{\tau_\textrm{c}}{3\tau_\textrm{2-Res.}}\right).
\end{split}
\label{Emean}
\end{equation}
\noindent The second term $\frac{1}{4}\Omega_\textrm{R}^2 \tau_\textrm{c}^2 \times k_\textrm{b}T_0\frac{\tau_\textrm{c}}{3\tau_{i-\textrm{Res.}}}$ is the effective energy transferred to one ion due to the heating effect on the other. When ion 1 is in thermal equilibrium $E_1(0)\equiv E_{1,\textrm{eq}}$ and $\Delta E_1(\tau_\textrm{c})=0$, its temperature $T_{1,\textrm{eq}}=E_{1,\textrm{eq}}/k_\textrm{b}$  can be calculated:
\begin{equation}
 T_{1,\textrm{eq}} \approx T_\textrm{D}+\frac{1}{3}T_0\frac{\tau_\textrm{c}}{\tau_\textrm{2-Res.}}+T_0\frac{4}{\Omega_\textrm{R}^2 \tau_\textrm{c} \tau_\textrm{1-Res.}}.
\label{Eeq}
\end{equation}

\noindent In the case of $\tau_\textrm{1-Res.} \gg \tau_\textrm{2-Res.}$, this expression has a minimal temperature $T_{1,\textrm{min}}$ of ion 1  with an optimal coupling time $\tau_{c,\textrm{opt}}$: 
\begin{equation}
\begin{split}
& T_{1,\textrm{min}} \approx T_0\frac{4}{\sqrt{3}\Omega_\textrm{R}}\sqrt{\frac{1}{\tau_\textrm{1-Res.} \tau_\textrm{2-Res.}}}+T_\textrm{D},\\
& \tau_{c,\textrm{opt}}=\frac{2}{\Omega_\textrm{R}}\sqrt{\frac{3\tau_\textrm{2-Res.}}{\tau_\textrm{1-Res.} - \tau_\textrm{2-Res.}}}.
\end{split}
\label{Tmin}
\end{equation}

\noindent The achievable temperature is always higher than $T_\textrm{D}$ due to noise heating of both ions. The excess term ($T_0\frac{4}{\sqrt{3}\Omega_R}\sqrt{\frac{1}{\tau_\textrm{1-Res.} \tau_\textrm{2-Res.}}} \propto T_0C_{\textrm{eff}}R_{\textrm{eff}}$) is the main limitation of this cooling method. For a high-$Q$ resonator or trap electrodes with low capacitance the limitation of the cooling temperature can be reduced down to the millikelvin regime. In the other case $\tau_\textrm{1-Res.} \ll \tau_\textrm{2-Res.}$, e.g. one HCI like $^{208}\textrm{Pb}^{81+}$ coupled with a few $^9\textrm{Be}^+$ ions, the equilibrium temperature takes the simple form:

\begin{equation}
 T_{1,\textrm{eq}} \approx T_\textrm{D} + T_0\frac{4}{\Omega_\textrm{R}^2 \tau_\textrm{c} \tau_\textrm{1-Res.}}.
\label{Teq2}
\end{equation}
\noindent Here, the excess term is only due to the heating of ion 1. 

\noindent At the beginning of the sympathetic cooling, ion 1 is at $T_1 \approx T_0 \gg T_{1,\textrm{eq}}$ while ion 2 can be laser cooled to $T_\textrm{D}$ in advance. The energy reduction of ion 1 in $\tau_c$ time is given by $\Delta E \approx -\frac{1}{4}k_\textrm{b}\Omega_\textrm{R}^2 \tau_c^2T_1$. From that we can expect the cooling to follow an exponential function: 
\begin{equation}
T_{1}(t)=(T_{0} - T_{1,\textrm{eq}})e^{-\frac{t}{\tau_{\textrm{eff}}}}+T_{1,\textrm{eq}}.
\label{E1t}
\end{equation}
\noindent Here, $\tau_{\textrm{eff}}=\frac{4}{\Omega_\textrm{R}^2 \tau_c}$ is the effective cooling time constant. If the optimized coupling time $\tau_{\textrm{c,opt}}$ is chosen in order to reach the minimal temperature, then $\tau_{\textrm{eff,opt}} \approx \frac{2}{\sqrt{3}\Omega_\textrm{R}}\sqrt{\frac{\tau_\textrm{1-Res.}}{\tau_\textrm{2-Res.}}} \propto \frac{D_1^2}{q_1^2}C_{\textrm{eff}}$. For a fast cooling, a small effective electrode distance $D$ and a small effective capacitance are favorable.

\begin{figure}[!h]
\centering
\includegraphics[width=0.45\columnwidth]{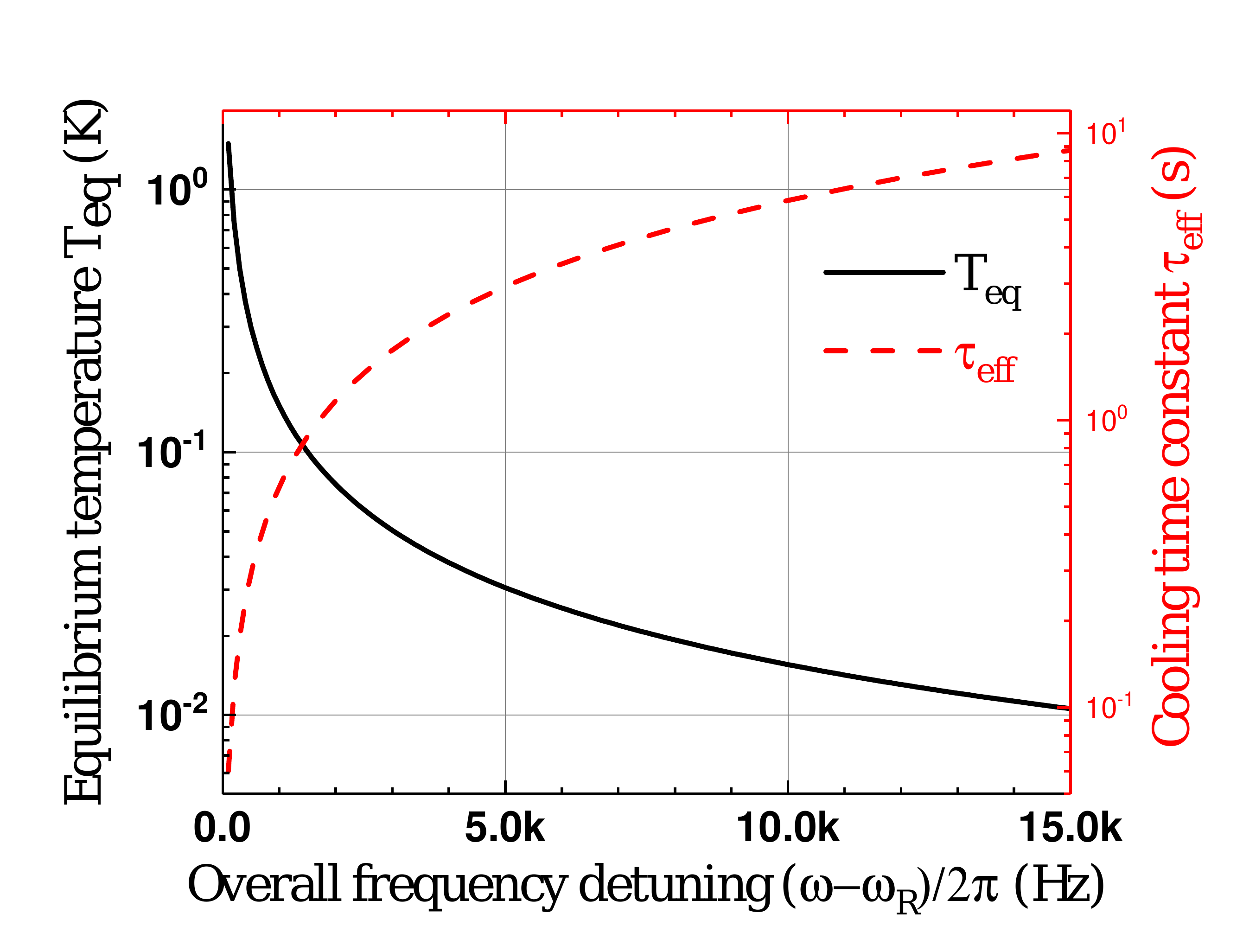}
\caption{The calculated equilibrium temperature $T_\textrm{eq}$ (solid black) and effective cooling time constant $\tau_\textrm{eff}$ (dashed red) of $\textsc{H}_2^+$ as a function of the overall frequency detuning $(\omega - \omega_\textrm{R})/2 \pi$ off the resonance frequency with the intermittent laser cooling method. For details see text.}
\label{Tf}
\end{figure}

\noindent We envisaged sympathetic cooling of a single $\textrm{H}_2^+$ ion with 100 $^9\textrm{Be}^+$ ions in the coupling trap (discussed in section \ref{comEC}, $D=4.6$ \SI{}{\milli\meter} and $C_\textrm{T} = 10$ \SI{}{\pico\farad}) assisted by the resonant tank circuit ($L = $ \SI{2.1}{\milli\henry}, $C_\textrm{R} = $ \SI{5.1}{\pico\farad}). Assuming a similar loss in the resonator $R_\textrm{p} = 344$ \SI{}{\mega\ohm} as currently in the PT \footnote {The loss in the resonator $R_\textrm{p}$ will change a little due to the higher resonance frequency from $2\pi \times 650$ \SI{}{\kilo\hertz} in PT to $2\pi \times 890$ \SI{}{\kilo\hertz} in the coupling trap, but it is hard to determine the result without testing.} the $Q$-value would be about 28,000. The equilibrium temperature (black solid) and the effective cooling time constant (red dashed) can be calculated with an optimized coupling time $\tau_{\textrm{c,opt}}$ as a function of the overall detuning $(\omega - \omega_\textrm{R})/2 \pi$ off the resonance frequency (shown in Figure \ref{Tf}). With larger detuning the final equilibrium temperature gets lower while the cooling becomes slower due to the increased effective capacitance (see Section \ref{prin}). If a larger coupling time length $\tau_{\textrm{c}}$ is used, the coupling becomes faster, however $T_\textrm{eq}$ is higher according to Equation (\ref{Eeq}). Using this scaling it is possible to design a cooling scheme that enables obtaining millikelvin equilibrium temperatures for a single $\textrm{H}_2^+$ ion within a reasonable time.

\begin{figure}[!h]
\centering
\includegraphics[width=0.45\columnwidth]{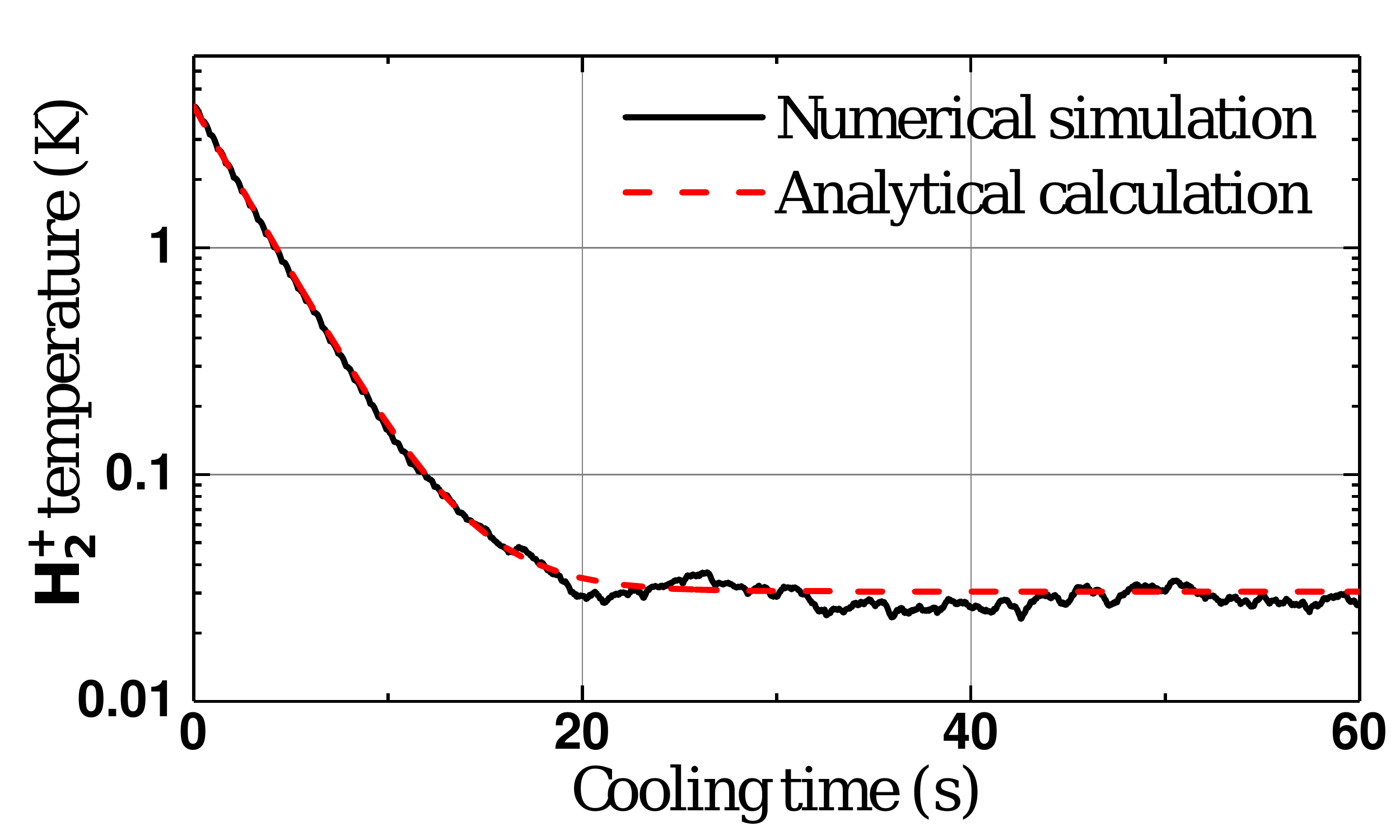}
\caption{The numerical simulation (black solid) and analytical calculation (red dashed) of sympathetic cooling of a single $\textsc{H}_2^+$ ion with 100 $^9\textrm{Be}^+$ ions assisted by a common tank circuit as a function of the cooling time.}
\label{sim}
\end{figure}

\noindent The sympathetic cooling can be also numerically simulated with Equation (\ref{RWA1}) where the damping terms due to the resonator coupling are $\gamma_{ii} = \frac{N_{i}q_{i}^2R_\textrm{eff}}{m_{i}D_{i}^2}$, $\gamma_{ij} = \frac{N_{j}q_{i}q_{j}R_\textrm{eff}}{m_{i}D_{i}D_{j}}$ and with the Johnson noise voltage $U_\textrm{noise} = \sqrt{4T_0k_bR_\textrm{eff}\Delta f}$ ($\Delta f $ is the bandwidth). We choose a coupling position \SI{5}{\kilo\hertz} off the resonance frequency to reduce the resonator heating. The optimized coupling length there is $\tau_{\textrm{c,opt}} \approx 0.4$ \SI{}{\second}. The simulation demonstrates that the $\textrm{H}_2^+$ ion is exponentially cooled down from \SI{4.2}{\kelvin} to about \SI{30}{\milli\kelvin} in about \SI{20}{\second}, which agrees with the analytical calculation from Equation (\ref{E1t}) (see Figure \ref{sim}). To achieve even lower temperatures, one can use either a larger detuning with a longer cooling time, increase the $Q$-value of the tank circuit or reduce the trap capacitance. Additionally, it is possible to adjust $\tau_{\textrm{c}}$ to the optimized value after a pre-cooling period with longer $\tau_{\textrm{c}}$ to achieve faster cooling and lower temperatures.

\begin{figure}[!h]
\centering
\includegraphics[width=0.45\columnwidth]{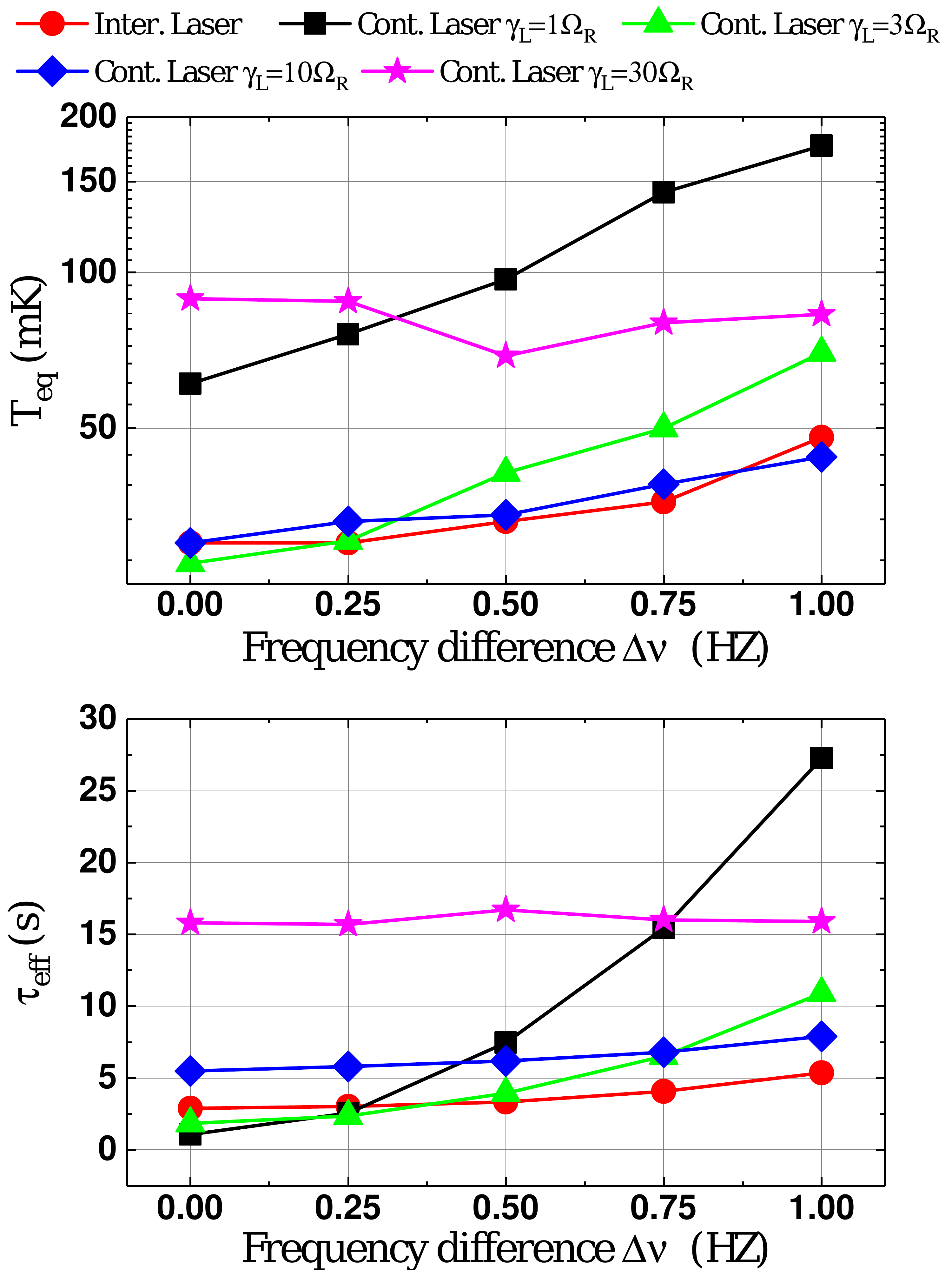}
\caption{The calculated equilibrium temperature $T_\textrm{eq}$ (upper plot) and effective cooling time constant $\tau_\textrm{eff}$ (lower plot) of $\textrm{H}_2^+$ as a function of the frequency difference $\Delta \nu$ with intermittent (red circle) and continuous laser (other colors and shapes) cooling methods. For details see text.}
\label{compr}
\end{figure}

\noindent For some precision measurements at \textsc{Alphatrap}, the target ion needs to stay in the PT, which has a spatial distance to the coupling trap. By connecting the same resonator to both traps, it is still possible to couple the target ion with the auxiliary ions for sympathetic cooling. As an example, a single $^{208}\textrm{Pb}^{81+}$ ion in the PT ($D_{PT} = 29.2$ \SI{}{\milli\meter} and $C_\text{T} = 23.3$ \SI{}{\pico\farad}) can be cooled to about \SI{20}{\milli\kelvin} in \SI{20}{\second} according to our simulation. 

\noindent Finally we can compare the intermittent and the continuous laser cooling technique. Simulations of $\textrm{H}_2$-$\textrm{Be}$ cooling with the same trap parameters have been done for both cooling methods, resulting in equilibrium temperatures and effective cooling time constants as a function of the frequency difference $\Delta \nu$ (shown in Figure \ref{compr}). With $\Delta \nu$ up to \SI{1}{\hertz}, $T_\textrm{eq}$ increases to \SI{48}{\milli\kelvin} and $\tau_\textrm{eff}$ increases by a factor of 2 for the intermittent laser cooling method. In some cases with a special laser damping coefficient $\gamma_\textrm{L}$ and small frequency mismatch, the continuous laser cooling method can achieve even lower temperatures in a reasonable cooling time. However, the intermittent laser cooling method is more robust with respect to the laser power and detuning and shows only small sensitivity to ion frequency stability as long as the coupling length $\tau_\textrm{c}$ is shorter than the inverse modified Rabi frequency $\Omega_{R}^{\prime}$, which can be always achieved by adjusting the frequency detuning $d\omega$.

\section{Conclusion}

In summary, a new technique for highly efficient sympathetic cooling has been proposed and tested in this work. With the assistance of a common resonator the axial motion of ion species located in separate traps can be strongly coupled. In the demonstration experiment, an avoided crossing behavior of the motion of $^{40}\textrm{Ar}^{11+}$ and $^{84}\textrm{Kr}^{23+}$ ions has been observed. In addition, an intermittent laser cooling method has been studied both in analytical calculations and numerical simulations for sympathetic cooling of a singly charged $\textrm{H}_2^+$ ion to \SI{30}{\milli\kelvin} in about \SI{20}{\second}. According to the technique presented in this paper, we show the possibility of cooling arbitrary types of ions to the millikelvin regime within reasonable cooling times.
% Experimental section

%\medskip
%\textbf{Supporting Information} \par %Please delete the Suppporting Information statement if it is not applicable. Please supply Supporting Information in another file. Supporting information should not be provided in .tex format

% Acknowledgements
\medskip
\textbf{Acknowledgements} \par %delete if not applicable))
We acknowledge financial support from the Max Planck Society. This work is supported by the German Research Foundation (DFG) Collaborative Research Centre SFB 1225 Project-ID 273811115 (ISOQUANT). This project received funding from the European Research Council (ERC) under the European Union’s Horizon 2020 research and innovation programme under grant agreement number 832848 - FunI. Furthermore, we acknowledge funding and support by the International Max Planck Research School for Quantum Dynamics (IMPRS-QD) and by the Max Planck, RIKEN, PTB Center for Time, Constants and Fundamental Symmetries. B.T. was supported by a Humboldt Research Fellowship for Postdoctoral Researchers. This Letter comprises parts of the Master thesis work of F. Hahne, Heidelberg University, Germany. The author thanks Prof. Klaus Blaum for his contributions to this project and the discussions and his help in the revision.

% References
\medskip

% Use the following code if you wish to generate your bibliography with BibTeX;
% replace the string "MSP-template" below with the name(s) of
% the BibTeX data base(s) you want to use.
% The resulting bibliography-output (the content of the .bbl file)
% must be pasted back into this file before submission.
% Please also include your BibTeX data base file(s) in your submission
% so that we can re-run BibTeX if necessary.
%
\bibliographystyle{MSP}
\bibliography{B_Tu_ref}

\end{document}